\documentstyle[12pt]{article}

\newcommand{\ee}{\end{equation}}
\newcommand{\be}{\begin{equation}}

\newcommand{\pa}{\partial}

 \newcommand{\r}[1]{(\ref{#1})}

\begin{document}
\thispagestyle{empty}
\vspace*{-1.5cm}
\begin{center}
{\LARGE\bf Dilute Instanton Gas of an O(3) Skyrme Model} \\[14mm] {\large
H.J.W. M\"uller--Kirsten\raisebox{0.8ex}{\small a,}\footnote{E--mail:
mueller1@physik.uni--kl.de},
S.--N. Tamarian\raisebox{0.8ex}{\small a,b,}\footnote{E--mail:
sayat@moon.yerphi.am},
\\
D.H. Tchrakian\raisebox{0.8ex}{\small c,d,}\footnote{E--mail:
tchrakian@may.ie} and
F. Zimmerschied\raisebox{0.8ex}{\small a,}\footnote{E--mail:
zimmers@physik.uni--kl.de}}
\\[1cm]
{\it \raisebox{0.8ex}{\small a} Department of Physics \\ University of
Kaiserslautern, P.\ O.\ Box 3049, D 67653 Kaiserslautern, Germany
\\[4mm] \raisebox{0.8ex}{\small b} Yerevan Physics Institute
\\ Alikhanian Br. Str. 2, 375 036 Yerevan 36, Armenia \\[4mm]
\raisebox{0.8ex}{\small c} Department of Mathematical Physics \\
National University of Ireland, Maynooth, Ireland
\\[4mm] \raisebox{0.8ex}{\small d}
School of Theoretical Physics, Dublin Institute for Advanced studies, \\ 10
Burlington Road, Dublin 4, Ireland} \\
\vfill
{\bf Abstract}
\end{center}

\noindent
The pure-Skyrme limit of a scale-breaking Skyrmed
$O(3)$ sigma model in 1+1 dimensions is employed to study the effect of the
Skyrme term on the semiclassical analysis of a field theory with instantons.
The instantons of this model are self-dual and can be evaluated explicitly.
They are also localised to an absolute scale, and their fluctuation action
can be reduced to a scalar subsystem. This permits the explicit calculation 
of the fluctuation determinant and the shift in vacuum energy due to instantons.
The model also illustrates the semiclassical quantisation of a Skyrmed field 
theory.
\\[0.6cm]
\newpage

\section{Introduction}

The physical motivation of the present work is to carry out the instanton 
analysis for a field theory described by a Lagrangian that includes 
Skyrme-like kinetic terms.
Such systems are expected to arise as low energy effective theories, and are
motivated by the physics at very high energy scale as described by superstring 
theory.
The semiclassical analysis of
physically interesting or relevant gauge field theories of this sort in 3+1 
dimensions is very difficult to handle, so instead we consider an ungauged 
system in 1+1 dimensions as a toy model. We choose to work with such a 
model that supports exponentially localised instantons which can be evaluated 
analytically. Furthermore the eigenvalue problem for the fluctuation operator 
can be solved in closed form. This enables us to evaluate the (regularised) 
determinants exactly and hence to find the shift in vacuum energy due to
instantons. Thus apart from affording us an example of the semiclassical
quantisation of a theory with a Skyrme
term, the model under consideration is remarkable in that it allows the complete
semiclassical (instanton) analysis of a field theory. We think that this feature
is of interest in its own right.


The model we employ, which was introduced in Ref.~\cite{MPTZ}, is a Skyrmed 
version of a scale-breaking $O(3)$ sigma model~\cite{MW}. 
While the latter system~\cite{MW} does not support finite-sized instantons, 
our model~\cite{MPTZ} does. These instantons are not self-dual and are 
evaluated only numerically. In the present work, where we need to take the 
analysis as far as possible, it is desirable that the instantons be explicitly 
evaluated. This leads us to consider what we call the pure-Skyrme limit
of the model in Ref.~\cite{MPTZ}, described by the Lagrangian
\begin{equation}
\label{1}
{\cal L} = \frac{\lambda_2}{4 \kappa^2} (\phi_{\mu \nu}^{ab})^2 +\lambda_1
(\phi_{\mu}^a)^2 + \lambda_0 \kappa^2 V(\phi^3) \end{equation} in the
$\lambda_1 \to 0$ limit. In \r{1} we have used the notation $\phi_{\mu}^a
=\partial_{\mu} \phi^a$, $\phi_{\mu \nu}^{ab} =\phi_{[\mu}^a \phi_{\nu
]}^b$, and of course the field $\phi$ is subject to $\phi^a \phi^a =1$,
with $a=1,2,3$. $\kappa$ is a constant with the dimension of inverse length,
and the coupling constants $\lambda_0 ,\lambda_1$ and $\lambda_2$ are
dimensionless.
The choice of the potential $V(\phi^3)$ which breaks the $O(3)$ 
symmetry to $SO(2)$ specifies the model.

The radially symmetric vorticity $N$ instanton solutions of this model were
found numerically in Ref.\ \cite{MPTZ}. Subsequently the self-dual
solutions of the pure-Skyrme limit with $\lambda_1 \to 0$ in \r{1} were
found analytically in Ref.\ \cite{ART}, where it was also shown that these
solutions were approximated by the numerically found solutions for small
values of $\lambda_1$, continuously. This property was established {\it
only for unit vorticity} $|N|=1$.

In the present work, where we propose to employ the analytic solutions of
the pure-Skyrme model
\begin{equation}
\label{2}
{\cal L} = \frac{\lambda_2}{4 \kappa^2} (\phi_{\mu \nu}^{ab})^2 +\lambda_0
\kappa^2 V(\phi^3)
\end{equation}
as instantons, we need to give some justifications. The problem is that the
vacuum field $\phi^a$ must be time independent, so in the context of our
Euclidean space formulation, it must also be $x_{\mu}$ independent. This is
assured by the vanishing of the quadratic kinetic term multiplying
$\lambda_1$ in \r{1} in the asymptotic (vacuum) region. (Note that the
vanishing of the quartic kinetic (Skyrme) term does not imply this.) Thus if
we are to employ the self-dual solutions to \r{2} as instantons, we
must consider the system \r{2} as an approximation to \r{1} with very small
$\lambda_1$. This is justified by the fact that the topologically stable
finite action solutions to \r{2} approximate those of \r{1} for very small
$\lambda_1$ \cite{ART}. The only restriction is that this situation holds
only for $|N|=1$ instantons, but given that the latter are {\it absolutely}
localised, it will be necessary to consider only unit topological charge
instantons anyway. This justifies our use of the pure-Skyrme model \r{2}
and its solutions as instantons, provided it is understood that \r{2} is
seen as an approximation of \r{1}.

Before we proceed, it may be relevant to note that in the limit of
latter, model \cite{BP} being scale invariant, supports instantons featuring
an arbitrary scale, while \r{1} and the model \r{2} employed here, are not
scale invariant and hence support instantons localised to an absolute
scale. Thus the instantons of both models \r{1} and \r{2} have {\it finite
size} so that there is no need for us to constrain the size of instantons
\cite{A} as is necessary when the
instantons are not fixed to an absolute scale. Another relevant remark is
that the model \r{2} does not support either sphalerons, as in Ref.\
\cite{MW}, or periodic instantons \cite{KRT}, unlike \r{1}. The periodic
instantons of \r{1} were found in Ref.\ \cite{KT}, while the sphaleron
induced thermal transitions in the model \r{1} were analysed in Ref.\
\cite{ZTM}. Here we will be concerned only with the (zero temperature) {\it
finite size} instantons of model \r{2}.

Accordingly, we have a 2 dimensional model at hand whose topologically
stable finite action solutions can be considered to be the instantons of
the theory. These instantons are known analytically, hence making the task
of calculating the fluctuation determinant tractable, although their
special symmetry properties result in an infinitely degenerate fluctuation
operator, yielding a system of only one scalar fluctuation field instead of
the two fluctuation parameter fields one expects for an $O(3)$--$\sigma$
model.

Furthermore, the localisation of these instantons to an absolute scale
prevents infrared
divergences arising from the integration over the scale collective
coordinate, with the result that it is possible to evaluate the vacuum
energy by employing a dilute gas of unit topological charge instantons.

In Section 2 we present the instanton solutions and expose their
symmetries. In Section 3 we derive the matrix fluctuation operator
around the instanton solution using a convenient parametrisation and
discuss the metric on the parameter space of fluctuations. In Section 4, we
calculate the normalised determinant, first discussing the 
infinite degeneracy of the fluctuation operator. After the unphysical degrees of
freedom related to the special symmetries of the model have been removed, we
determine the spectrum of the resulting scalar fluctuation operator using
zeta function regularisation \cite{H,S} and rescaling normalisation.
Finally, we calculate the shift in vacuum energy due to a dilute instanton gas.

\section{The instantons}

The analytic evaluation of the instantons will be carried out for the family of
models \r{2} specified by the potentials
\begin{equation}
\label{3}
V(\phi^3)=(1-\phi^3)^{k} , \qquad k=2,3,\ldots .
\end{equation}
For the case $k=1$ there does not exist a solution which is
analytic at the origin, hence $k=1$ in \r{3} is excluded.

The equations of motion of the models \r{2}--\r{3} are satisfied by the
solutions of the following Bogomol'nyi equations \begin{equation}
\label{4}
\sqrt{\lambda_2}\varepsilon_{\mu \nu} \varepsilon^{abc} \partial_{\mu}
\phi^a \partial_{\nu} \phi^b \phi^c -
\sqrt{\lambda_0}(1-\phi^3)^{\frac{k}{2}} = 0 \end{equation} The topological
charges, which are equal to the action, are given by the volume integrals
of the densities
\begin{equation}
\label{5}
\varrho = 2\sqrt{\lambda_2\lambda_0}
\varepsilon_{\mu \nu} \varepsilon^{abc} \partial_{\mu} \phi^a
\partial_{\nu} \phi^b \phi^c(1-\phi^3)^{\frac{k}{2}} . \end{equation}

Before proceeding to solve \r{4}, we note that the solutions of the
self-duality equation \r{4} satisfy an infinite dimensional symmetry, which
will be treated as an unphysical symmetry in Section 4.
This is in contrast with the physical, finite dimensional, symmetries of \r{2}
under spacetime translations and rotations. It is straightforward to check that
\r{4} is invariant under
\begin{equation}
\label{5a}
\phi^{\alpha} \rightarrow R^{\alpha}{}_{\beta}\phi^{\beta} \qquad \phi^3
\rightarrow \phi^3 ,
\end{equation}
in which the $SO(2)$ rotation
$R^{\alpha}{}_{\beta}=R^{\alpha}{}_{\beta}(|\phi^{\gamma}|)$ is a function of
the modulus $|\phi^{\alpha}|$ of the functions $\phi^{\alpha}$, i.e.\ there
is a local rotation symmetry in the inner symmetry space besides the usual
global $O(3)\rightarrow SO(2)$ broken symmetry of the model which is equivalent
to spacetime rotations. In Section 4, we shall find that this local symmetry
results in a degenerate fluctuation operator around the solutions of
equation \r{4}.

We now proceed to solve \r{4} and to that end consider the radially
symmetric field configuration of vorticity $N$ \begin{equation} \label{6}
\phi^{\alpha} =\sin f(\rho)\: n^{\alpha} , \qquad \phi^3 =\cos f(\rho)
\end{equation}
where we have used the dimensionless radial variable $\rho =\kappa r$
defined in terms of $\kappa$ appearing in \r{1}--\r{2}, and where
$n^{\alpha} =(\cos N\theta , \sin N\theta)$ is a unit vector with winding
number $N$.
The Euler-Lagrange equations of this system for a particular choice of the
potential $V(\cos f)$ were found in \cite{ART}, subject to the asymptotic
conditions
\begin{equation}
\label{7}
\lim_{\rho \rightarrow 0} f(\rho) =\pi ,\qquad \lim_{\rho \rightarrow
\infty} f(\rho)=0. \end{equation}
In the radially symmetric field configuration \r{6}, the reduced
Bogomol'nyi (self-- or antiself--dual) equation and the topological 
charge density of the resulting
1-dimensional subsystem corresponding to (\ref{4}) and (\ref{5})
respectively are
\begin{equation}
\label{8}
\frac{N}{\rho}\sin f \frac{df}{d\rho} = \pm\beta(1-\cos f)^{\frac{k}{2}}
\quad \mbox{with} \quad
\beta=\sqrt{\frac{\lambda_0}{\lambda_2}} \end{equation} \begin{equation}
\label{9}
\varsigma =4\pi N\sqrt{\lambda_2\lambda_0} (1-\cos f)^{\frac{k}{2}} \sin f
\frac{df}{d\rho} .
\end{equation}
For $N>0$ we get solutions satisfying the boundary conditions (\ref{7})
only if we choose the negative sign in \r{8}. We call these solutions
instantons although they solve the antiself--dual equation.
The antiinstantons then, with $N<0$, are the solutions of the self--dual
equation with the positive sign in \r{8}. Instantons and antiinstantons
hence differ in the direction of the vorticity $n^{\alpha}$ \r{6}, but not 
in their radial behaviour $f(\rho)$.

The resulting solutions are 
\begin{eqnarray}
\label{10}
f(\rho) & = & \arccos\left(1-2e^{-\frac{\beta}{2|N|}\rho^2}\right), 
\qquad k=2 \\
\label{11}
f(\rho) & = & \arccos\left(1-\left[\frac{\beta}{2|N|} \frac{k-
2}{2}\rho^2+2^{\frac{2-k}{2}}\right]^{\frac{2}{2-k}}\right), \qquad k>2
\end{eqnarray}
satisfying the asymptotic conditions (\ref{7}) as required, and resulting
in the topological charges $N$ times the normalisation factor appearing in
(\ref{9}) multiplied further by $\frac{2^{\frac{k}{2} +1}}{\frac{k}{2}+1}$
for each $k>2$.

The functions \r{10}--\r{11} will be genuine solutions to the full
equations \r{4} only if they describe fields $\phi^a$ that are
singlevalued at the origin. It is easy to find the behaviour of $f(\rho)$
in the region $\rho \ll 1$ for the solutions (\ref{10})--(\ref{11}). The
conclusion is that for all these solutions we have the following behaviour
for $\sin f(\rho)$ \begin{equation}
\label{12}
\sin f \sim \rho, \qquad \qquad \rho \ll 1. \end{equation}
This means that the solution field $\phi^a$ given by (\ref{6}) is
singlevalued at the origin {\it only for vorticity} $|N|=1$. We must
therefore reject all solutions (\ref{10})--(\ref{11}) {\it except those of
unit vorticity}. This is perfectly consistent with our intention of
considering a dilute gas of widely separated instantons and anti-instantons
of {\it unit topological charge}, localised to an absolute scale $\kappa
=\rho /r$.

We shall restrict henceforth to the $N=1$ instanton of the model \r{2}
characterised by the potential \r{3} with $k=2$. The latter choice is made
because the instantons \r{10} of that model are localised {\it exponentially}
rather than by a {\it power} decay as in \r{11}. The action of this instanton
is readily calculated to be
\begin{equation}
\label{action}
S_0(\lambda_2,\lambda_0)=\frac{8\pi}{\beta}\lambda_0. \end{equation}

\section{Fluctuations around the instanton and the vacuum--vacuum transition
rate}

To evaluate the contributions of quantum fluctuations around the background
instanton field
\r{10},
we find it convenient to use a parametrisation of the fields $\phi^a$ which
satisfies the constraint $\phi^a\phi^a =1$ automatically, so that we do not
have to take this constraint into account by introducing the Lagrange
multiplier implicit in \r{1} and \r{2}. Note that it would have been
necessary to employ the Lagrange multiplier had we tried to solve the
Euler-Lagrange equations of \r{2}, but it was not necessary to do so when
we solved the self-duality equation \r{4}.

Such a parametrisation is
\begin{equation}
\label{12a}
\phi^a =\left(
\begin{array}{c}
\phi^{\alpha} \\
\phi^3
\end{array}
\right)
=\left(
\begin{array}{c}
\sin \Theta \cos \Phi \\
\sin \Theta \sin \Phi \\
\cos \Theta
\end{array}
\right)
\end{equation}
in terms of which the Lagrangian \r{2} specified by the potential \r{3}
with $k=2$ is expressed as
\begin{equation}
\label{13}
{\cal L} = \frac{\lambda_2}{4 \kappa^2} |\pa_{[\mu}\Theta \pa_{\nu ]}
\Phi|^2 \sin^2 \Theta
+\lambda_0 \kappa^2 (1-\cos \Theta)^2 .
\end{equation}
The instanton solution in whose background we calculate the second
variation of \r{13} is
\begin{equation}
\label{14}
\Theta = f(\rho) , \qquad \Phi =\theta,
\end{equation}
where $f(\rho)$ is given by \r{10}.

The Lagrangian \r{13} has all the invariances of \r{2}, namely spacetime
translations and rotations, as well as the infinite dimensional
local rotation symmetry which in terms of the parameter fields is given by
$\theta$ dependent translations of the field $\Phi$. Whereas the former
spacetime symmetries are broken by the solution \r{14}, the Bogomol'nyi
equation \r{8} and hence the solution itself is invariant under the latter
local symmetry.

Fluctuations $\delta\phi^a$ around a configuration $\phi^a$ are related to
fluctuations $\delta \Theta$, $\delta \Phi$ of the parameter fields by
\begin{equation}
\label{f01}
\delta\phi^a = e_{\Theta}^a(\Theta,\Phi)\delta \Theta +
e_{\Phi}^a(\Theta,\Phi)\delta \Phi
\end{equation}
with
\begin{equation}
\label{f02}
e_{\Theta}^a(\Theta,\Phi):=\left(\begin{array}{c} \cos \Theta \cos \Phi \\
\cos \Theta \sin \Phi \\
-\sin \Theta
\end{array}\right), \quad
e_{\Phi}^a(\Theta,\Phi):=
\left(\begin{array}{c}
-\sin \Theta \sin \Phi \\
+\sin \Theta \cos \Phi \\
0
\end{array}\right).
\end{equation}
Such fluctuations are elements of the tangent space to the inner symmetry
space ${\rm S}^2$ of the $O(3)$--$\sigma$ model, $\delta\phi^a\in {\rm
T}_{\phi^a}{\rm S}^2$. The parametrisation (\ref{12a}) induces a basis in
this tangent space.
With respect to this basis, the inner
product of parameter tangent space vectors $\vec{\xi}_i$ is given by
$\langle\vec{\xi}_1,\vec{\xi}_2\rangle=\vec{\xi}_1^T\hat{{\cal
G}}\vec{\xi}_2$ with the induced metric
\begin{equation}
\label{f0}
\hat{{\cal G}} = \left(\begin{array}{c}
e^a_{\Theta}(f,\theta)\\
e^a_{\Phi}(f,\theta) \end{array}\right)
\left(e^a_{\Theta}(f,\theta),e^a_{\Phi}(f,\theta) \right) =
\left(\begin{array}{cc} 1 & 0 \\ 0 & \sin^2 f \end{array} \right)
\end{equation}
The Skyrme term in the parameter field Lagrangian (\ref{13}) is related to
this inner product since it may be written as the square of the totally
antisymmetric wedge product defined on the basis of this inner product.

It is now laborious but straightforward to calculate the second variation
$\delta^{(2)}S$ of the action in the background of \r{14} which is the
quadratic fluctuation contribution around the classical instanton solution.
Denoting the variations of $\Theta$ and $\Phi$ by
$\delta \Theta=u(\rho ,\theta)$ and $\delta \Phi =v(\rho
,\theta)$, we find \begin{equation}
\label{f1}
\delta^{(2)}S[u,v] = \int
(u,v)\hat{{\cal M}}{u \choose v} \rho d\rho d\theta \end{equation} with the
$2\times 2$ matrix differential operator \begin{equation} \label{f2}
\hat{{\cal M}} := \left(
\begin{array}{cc}
\lambda_2\hat{L}^{\dagger}(1-\cos f(\rho))^2\hat{L} & -\sqrt{\lambda_0\lambda_2}
\hat{L}^{\dagger}(1-\cos f(\rho))^2\frac{\partial}{\partial\theta}\\
\sqrt{\lambda_0\lambda_2}(1-\cos f(\rho))^2\hat{L}
\frac{\partial}{\partial\theta} & -\lambda_0(1-\cos f(\rho))^2
\frac{\partial^2}{\partial\theta^2}
\end{array}\right)
\end{equation}
where $\hat{L}$ is a first--order differential operator,
\begin{equation}
\label{f3}
\hat{L}:=\frac{1}{\rho}\frac{\partial}{\partial \rho} \frac{\sin
f(\rho)}{1-\cos f(\rho)}, \qquad \hat{L}^{\dagger}= -\frac{\sin
f(\rho)}{1-\cos f(\rho)} \frac{1}{\rho}\frac{\partial}{\partial \rho}.
\end{equation}
The dagger denotes the adjoint in the corresponding Hilbert space. In
particular, $\hat{{\cal M}}^{\dagger}=\hat{{\cal M}}$ is self--adjoint in
the parameter space of the fluctuations with respect to the usual inner
product which appears in eq.\ (\ref{f1}).

To calculate the one instanton vacuum--vacuum transition rate, one has to
insert the fluctuation expansion of the Euclidean action around the instanton
solution into the path integral,
\begin{eqnarray}
\label{f4}
\langle 0| 0_V\rangle_1(\lambda_2,\lambda_0) & = & 
\int_{|0\rangle\stackrel{1 Inst}{\rightarrow}|0_V\rangle} 
{\cal D}\{\phi\} e^{-S[\phi]} \nonumber \\
& = & e^{-S_0(\lambda_2,\lambda_0)} 
\int {\cal D}\{u\}{\cal D}\{v\} e^{-\delta^{(2)}S[u,v]} =
\frac{e^{-S_0(\lambda_2,\lambda_0)}}{\sqrt{\det \hat{{\cal M}}}}
\end{eqnarray}
where the Euclidean fluctuation path integral results in the determinant of 
$\hat{{\cal M}}$ in the usual way. $V$ denotes the Euclidean spacetime
volume of the instanton interpolating between the two vacuum states.

Substituting the fluctuation parameter fields $u$, $v$ for the original
field $\phi$ one has to take into account the symmetry properties 
of the parameter field measure ${\cal D}\{u\}{\cal D}\{v\}$ which is related 
to the induced metric $\hat{{\cal G}}$ on the space of the fluctuation fields. 
Hence expanding the fluctuations $(u,v)$ in terms of the eigenfunctions
$\vec{\xi}_i$ of $\hat{{\cal M}}$ with respect to the measure $\hat{{\cal
G}}$, the determinant could be calculated as the product of the eigenvalues
in the equation \begin{equation}
\label{f5}
\hat{{\cal M}}\vec{\xi}_i = \tilde{\omega}^2_i \hat{{\cal G}} \vec{\xi}_i.
\end{equation}
The corresponding eigenfunctions can be orthonormalised with respect to the
metric $\hat{{\cal G}}$.

In general, the eigenvalue problem \r{f5} is difficult to solve. The only
eigenfunctions which are easy to find are the physical zero modes related to 
the translational and rotational symmetries of the Lagrangian broken by the
instanton solution $(\Theta,\Phi)=(f(r),\theta)$. In terms of this solution,
the physical zero modes normalised with respect to the metric $\hat{\cal G}$
are given by
\begin{equation}
\label{g1}
\vec{\xi}_x=
\frac{1}{\kappa\sqrt{c}}\frac{\partial}{\partial x} \left(\begin{array}{c}
f(r) \\ \theta
\end{array}\right),\: \:
\vec{\xi}_y=
\frac{1}{\kappa\sqrt{c}}\frac{\partial}{\partial y}\left(\begin{array}{c}
f(r) \\ \theta\end{array}\right),\: \:
\vec{\xi}_{\theta}=\sqrt{\frac{\beta}{4\pi}}\frac{\partial}{\partial
\theta}\left(\begin{array}{c} f(r) \\ \theta \end{array}\right)
\end{equation}
where
\begin{equation}
\label{g2}
c=4\pi\left(\frac{\pi^2}{12}+\log 2\right), \qquad
\beta=\sqrt{\frac{\lambda_0}{\lambda_2}}. 
\end{equation}
These zero modes correspond to the fact that the instanton can be localised at
an arbitrary spacetime point $x_0,y_0\in V$ with angular direction
$\theta_0\in(0,2\pi)$.

\section{Evaluation of the fluctuation determinant: factorisation and
regularisation}

Instead of solving the eigenvalue equation \r{f5} for the complete spectrum
and calculating the determinant as product of all eigenvalues, we now
proceed in a different way, exploiting the important feature that the matrix 
operator $\hat{{\cal M}}$ can be factorised, 
\begin{equation}
\label{f7}
\hat{{\cal M}} = \hat{{\cal N}}^{\dagger}\hat{{\cal N}}, \qquad \hat{{\cal
N}} = \left(1-\cos f(\rho)\right)\left( \begin{array}{cc} \sqrt{\lambda_2}
\hat{L} & -
\sqrt{\lambda_0} \frac{\partial}{\partial \theta} \\ 0 & 0 \end{array}
\right). 
\end{equation}
It is easy to check that $\hat{{\cal G}}\hat{{\cal N}}=\hat{{\cal N}}$,
hence we can evaluate $\det \hat{{\cal M}}$ as product of the eigenvalues
of the equation
\begin{equation}
\label{f9}
\hat{{\cal M}}\vec{\chi}_i = \omega^2_i \vec{\chi}_i. \end{equation}
Therefore the metric $\hat{{\cal G}}$ enters the final result only through
the normalisation of the zero modes of $\hat{{\cal M}}$.

When the eigenvalue equation \r{f9} is acted on from the left by the
operator $\hat{{\cal N}}$ defined by \r{f7}, the resulting eigenvalue
equation (for the new eigenfunction) has the same eigenvalues, now, however, 
with eigenfunctions
$\vec{\psi}_i=\hat{{\cal N}}\vec{\chi}_i$. Since we are only interested in
evaluating the
determinant, i.e. in the eigenvalues but not in the eigenfunctions, we choose
to work with the second eigenvalue equation \be
\label{f10}
\hat{{\cal N}}\hat{{\cal N}}^{\dagger} \psi_n =\omega_n^2 \psi_n . \ee
The new eigenvalue equation \r{f10} has a much simpler structure than 
the previous
one \r{f9}. This is because the operator \begin{equation}
\hat{{\cal N}}\hat{{\cal N}}^{\dagger}=
(1-\cos f)\left(\begin{array}{cc} \hat{H} & 0 \\ 0 & 0 \end{array}\right)
(1-\cos f)
\end{equation}
with
\be
\label{f12}
\hat H =\lambda_2 \hat L \hat L^{\dagger} -\lambda_0 \frac{\pa^2}{\pa
\theta^2}, \ee
($\hat L$ and $\hat L^{\dagger}$ given by \r{f3}) is diagonal, and even
more importantly, has only one nonvanishing diagonal element. This means the 
operator $\hat{{\cal N}}\hat{{\cal N}}^{\dagger}$ is infinitely degenerate due 
to the area (2 dimensional volume) preserving coordinate transformations 
resulting in the physical zero modes \r{g1}, and the transformations \r{5a}. 

Hence for every eigenvalue we have an eigenspace rather than an
eigenvalue, each element of this space arising from the action of the
elements of the infinite dimensional symmetries on a definite eigenfunction
in that eigenspace. We shall keep only one function from every such
eigenspace, which amounts to the elimination of unphysical degrees of
freedom. This amounts to rejecting the zero diagonal element
in $\hat{{\cal N}}\hat{{\cal N}}^{\dagger}$, with the result that the
determinant does not vanish. 

As usual we shall treat the translational and
rotational symmetries as physical, and will take their zero-modes
into consideration, integrating over the collective cooridnates $x_0$, $y_0$ 
and $\theta_0$ later on. Here, we exclude also these last
zero eigenvalues from the operator $\hat{{\cal N}}\hat{{\cal
N}}^{\dagger}$, and denote the
resulting determinant by $\det' \hat{{\cal M}}$, 
\be
\label{f11}
{\rm det}'\hat{{\cal M}} ={\rm det} \left[(1-\cos f)\hat H (1-\cos f)\right]. 
\ee
In fact, this reduces the fluctuation analysis from two parameter
fluctuation fields to only one scalar fluctuation field and hence
simplifies the analysis considerably.

Next, there arises the question of the normalisation of the determinant.
Since our theory \r{2} does not have a perturbative vacuum, we are forced
to choose an arbitrary normalisation point. In terms of the two
dimensionless coupling constants $(\lambda_0 ,\lambda_2)$, we choose this
point to be $(\lambda_0' ,\lambda_2')$ satisfying
\be
\label{f13}
\frac{\lambda_0'}{\lambda_2'} =\frac{\lambda_0}{\lambda_2}, 
\ee
amounting to a rescaling of the Lagrangian \r{2}, but keeping
$\beta^2=\lambda_0/\lambda_2$ fixed. As a result, the $\beta$--dependent
divergent contribution of $\det(1-\cos f)^2$ to $\det' \hat{{\cal M}}$
cancels in the normalised one instanton vacuum--vacuum transition rate
\begin{equation}
\label{zzzz}
\frac{\langle 0|0_V\rangle_1 (\lambda_2,\lambda_0)}
{\langle 0|0_V\rangle_1({\lambda}^{\prime}_2,{\lambda}^{\prime}_0)}
=e^{-[S_0(\lambda_2,\lambda_0)-S_0(\lambda_2',\lambda_0')]}
\left[\frac{\det' \hat{{\cal M}}(\lambda_2,\lambda_0)}
{\det' \hat{{\cal M}}(\lambda_2',\lambda_0')}\right]^{-\frac12}{\cal Z}
\end{equation}
where ${\cal Z}$ denotes the contribution of the zero modes.

Hence the evaluation of $\det'\hat{{\cal M}}$ reduces further to the evaluation
of the product of the eigenvalues of the following eigenvalue equation 
\[
\hat H \psi =\omega^2 \psi
\]
or
\be
\label{f14}
\bigg( -\lambda_2 {1\over \rho}\frac{\pa}{\pa \rho} \frac{1+\cos f}{1-\cos
f} {1\over \rho}\frac{\pa}{\pa \rho} -\lambda_0 \frac{\pa^2}{\pa
\theta^2}\bigg) \psi = \frac{\omega_n^2}{\kappa^2} \psi .
\ee
Using the notation $z=\cos f(\rho)$ and looking for separable solutions
$\psi (z, \theta) ={\rm e}^{im\theta} P(z)$, with $m=0,1,2,...$, the
eigenvalue equation \r{f14} reduces to the ordinary diferential equation
\be
\label{f15}
\lambda_0 \kappa^2 (1-z){d\over dz}(1+z){d\over dz}P(z) =(\omega^2
-\lambda_0 \kappa^2 m^2)P(z).
\ee
The solutions of this equation are the Jacobi polynomials, yielding the
eigenvalue spectrum
\be
\label{f16}
\omega_{n,m}^2 =\lambda_0 \kappa^2 (n^2 +m^2), \ee
with the integer $n=1,2,3,\ldots$.

Since the spectrum of the operator $\hat{H}$ is discrete, we will use the
method of Zeta function regularisation to evaluate its determinant. The
generalised Zeta function for this operator is 
\be
\label{f17}
\zeta_{\hat{H}} (s) =\sum_{n,m} \omega_{n,m}^{-s} . \ee
This series converges for Re$s \ge {5\over 2}$ (see \r{f22} below) and can
be analytically continued to a meromorphic function of $s$ which has no
singularity at $s=0$. This allows us to define the determinant as follows
\be
\label{f18}
-{1\over 2} {\rm log \: det} \hat H ={d\over ds}\zeta_{\hat{H}} (s)|_{s=0}. 
\ee
The summation over $m$ in \r{f17} yields \begin{equation}
\label{f19}
\zeta_{\hat{H}} (s) = ( \lambda_0 \kappa^2 )^{- \frac{s}{2}} \frac{2 \pi
}{\Gamma (\frac{s}{2})} \sum^{\infty}_{n = 1} (2 n)^{\frac{1 -
s}{2}}\int^{\infty}_{0} \frac{t^{\frac{s}{2} -1}}{e^t -1} J_{\frac{s-1}{2}}
(nt)dt + (\lambda_0 \kappa^2)^{- \frac{s}{2}} \zeta (s) , \end{equation}
in which $J_{\frac{s-1}{2}} (nt)$ is a Bessel function and $\zeta (s)$ is
the usual Zeta
function. After summing over $n$ in \r{f19}, we have \be
\label{f20}
\zeta_{\hat{H}} (s) = ( \lambda_0 \kappa^2 )^{- \frac{s}{2}}[ 2h_1(s) + 2
h_2(s) + 2h_3(s) + \zeta(s) ],
\end{equation}
where
\begin{eqnarray}
\label{f21}
h_1(s) &=& \sqrt{\pi} \frac{2^{- s}
}{\Gamma(\frac{s}{2})\Gamma(\frac{s+1}{2})} \int^{\infty}_{0} \frac{t^{s -
\frac{3}{2}}}{e^t -1} dt \\ \label{f22}
h_2(s) &=& 2 {\pi} \frac{2^{- s} }{\Gamma^2 (\frac{s}{2})}\int^{\infty}_{0}
\frac{t^{s - \frac{5}{2}}}{e^t -1} dt \\ \label{f23}
h_3(s) &=& 2 {\pi} \frac{2^{- s} }{\Gamma^2 (\frac{s}{2})} \sum^{\infty}_{n
= 1} \int^{2 \pi (n +1) }_{2 \pi n} \frac{(t^2 - 4 \pi^2 n^2)^{\frac{s}{2}
-1 }}{e^t - 1}\frac{dt}{\sqrt{t}}. \end{eqnarray}
The function $\zeta_{\hat{H}}(s)$ comprises radial fluctuation $\zeta(s)$,
and angular fluctuation
$ h_i(s)$, $i = 1,2,3$, contributions.
Since $ \zeta^{\prime}_{\hat{H}}(0) $ contains only $ h_i(0) $ and $
h^{\prime}_{i}(0)$ terms, we will keep only constant terms and terms linear
in $s$ in $ h_i(s) $. It is easy to see that $ h_2(s) $ has no such terms and
so in the limit $ s \rightarrow 0 $ we obtain
\begin{eqnarray}
\label{f24}
h_1(s) &\approx &- \sqrt{\pi} \zeta\left(- \frac{1}{2}\right) s \\ \label{f25}
h_3(s) &\approx &\frac{1}{2} \frac{h_0}{\sqrt{2 \pi}} s \end{eqnarray}
where in \r{f25} $h_0$ is given by
\begin{equation}
h_0 = \sum^{\infty}_{n =1 } \frac{n^{- \frac{3}{2}}}{e^{2 \pi n} -1 } =
0.00187217 .
\end{equation}
Because $ h_i(0) = 0 $, it follows that the angular fluctuation
contributions to the determinant
amount to a numerical factor only, while the parametric dependence is
controlled by the contributions of the radial fluctuations.

The result of the foregoing analysis is that \begin{eqnarray}
\zeta^{\prime}_{\hat{H}}(0) &=& \frac{1}{4} {\rm log}(\lambda_0 \kappa^2 )
- \frac{1}{2} {\rm log}(2 \pi ) -2 \sqrt{\pi} \zeta\left(
-\frac{1}{2}\right) + \frac{h_0}{\sqrt{2 \pi}} \nonumber \\
&=& \frac{1}{4} {\rm log}(\lambda_0 \kappa^2 ) - 0.181254 , \end{eqnarray}
and accordingly
\begin{equation}
\label{determinant}
\left[\det \hat{H}(\lambda_0)\right]^{- \frac{1}{2} } = {\rm exp}\left[
\zeta^{\prime}_{\hat{H}}(0)
\right]
= 0.8344223(\lambda_0 \kappa^2)^{\frac{1}{4}}. 
\end{equation}

Finally, we have to take into account the zero modes, i.e.\ to integrate 
along the directions in which the action does not change in the path
integral \r{f4}. These directions are just those of the collective
coordinates, but 
the corresponding measures $dc_x$, $dc_y$, $dc_{\theta}$ involve
the normalisation \r{g2} of the zero modes (this is where the metric
$\hat{{\cal G}}$ enters the calculation),
\begin{equation}
\label{g4}
\frac{dc_x}{\sqrt{2\pi}}=\kappa\sqrt{c}\frac{dx_0}{\sqrt{2\pi}},\: \:
\frac{dc_y}{\sqrt{2\pi}}=\kappa\sqrt{c}\frac{dy_0}{\sqrt{2\pi}},\: \:
\frac{dc_{\theta}}{\sqrt{2\pi}}=\sqrt{\frac{4\pi}{\beta}}
\frac{d\theta_0}{\sqrt{2\pi}}.
\end{equation}
The zero mode factor ${\cal Z}$ in eq.\ \r{zzzz} thus reads
\begin{equation}
\label{zero}
{\cal Z}=\frac{2\kappa^2c}{\pi\sqrt{2\beta}}\int dx_0dy_0d\theta_0
\end{equation}
where the $\theta$--integration contributes a factor  
$2\pi$ and the $x_0y_0$--integration yields the instanton volume $V$.

\section{Instanton density and vacuum energy shift}

Collecting the results \r{action},
\r{determinant}, \r{zero} of the previous sections, 
the one instanton contribution 
to the normalised vacuum--vacuum transition amplitude \r{zzzz} reads
\begin{equation}
\label{g5}
\frac{\langle 0|0_V\rangle_1 (\lambda_2,\lambda_0)}{\langle 0|0_V
\rangle_1({\lambda}^{\prime}_2,{\lambda}^{\prime}_0)}= {\rm
e}^{-[S_0(\lambda_2,\lambda_0)
-S_0({\lambda}^{\prime}_2,{\lambda}^{\prime}_0)]} \left[\frac{\det
\hat{H}(\lambda_0)}{\det \hat{H}({\lambda}^{\prime}_0)} \right]^{-\frac{1}{2}}
\frac{4\kappa^2c}{\sqrt{2\beta}}V = RV, 
\end{equation}
where the quantity
\begin{equation}
\label{g7}
R=2\sqrt{2}\frac{\kappa^2c}{\sqrt{\beta}}
\left(\frac{\lambda_0}{\lambda^{\prime}_0}
\right)^{\frac{1}{4}}e^{-\frac{8\pi}{\beta}(\lambda_0-{\lambda}^{\prime}_0)}
\end{equation}
is recognised as the instanton density. 
>From this expression for $R$ we deduce
the condition for diluteness of the instanton gas, i.e.\ 
\begin{equation}
\label{g8}
\frac{8\pi}{\beta}(\lambda_0-{\lambda}^{\prime}_0)\gg1 . \end{equation}

In the region of validity of the dilute gas approximation, the $n$
instanton contribution will be
\begin{equation}
\label{g9}
\frac{\langle 0|0_V\rangle_n (\lambda_2,\lambda_0)}{\langle 0|0_V
\rangle_n({\lambda}^{\prime}_2,{\lambda}^{\prime}_0)} =
\frac{1}{n!}\left(RV\right)^n . \end{equation}
Then the degree of suppression of the instanton generated transition
amplitude is obtained by summing over $n$,
\begin{equation}
\label{g10}
\frac{\langle 0|0_V\rangle (\lambda_2,\lambda_0)}{\langle 0|0_V
\rangle ({\lambda}^{\prime}_2,{\lambda}^{\prime}_0)}= \sum^{\infty}_{n = 1}
\frac{1}{n!}\left(RV\right)^n= e^{RV}.
\end{equation}
In the limit $V\rightarrow \infty$ we obtain \begin{equation}
\label{g11}
\langle 0|0_V\rangle \approx e^{-EV}
\end{equation}
where $V$ is the space volume, and hence $E$ can be treated as the vacuum
energy density. Finally, substitution of the expression \r{g11} in \r{g10}
yields the lowering shift in vacuum energy density due to the
instantons (as evidenced by the dependence on $\lambda_2$)
\begin{equation}
\label{g12}
E(\lambda_2,\lambda_0)-E(\lambda^{\prime}_2,\lambda^{\prime}_0)=-8\pi
\sqrt{2}\left(\frac{\pi^2}{12}+\log 2\right)\frac{\kappa^2}{\sqrt{\beta}}
\left(\frac{\lambda_0}{\lambda^{\prime}_0}\right)^{\frac{1}{4}}
e^{-\frac{8\pi}{\beta}(\lambda_0-{\lambda}^{\prime}_0)} \end{equation}
whose region of validity is controlled by condition \r{g8}.

\section{Summary}

In the above we performed and explicit evaluation of  
the normalised and regularised determinant of the
fluctuation operator
for an $O(3)$ Skyrme model in 2 dimensions. This calculation yielded the
exact expressions for the dilute instanton gas density and the 
corresponding shift in vacuum
energy. All this was made possible by three salient properties of the model
used: first the fact that the instantons of this model were explicitly
found, and second the special symmetry properties of the model which reduced
the fluctuation action to a scalar system, both these features 
enabling the exact calculation of
the fluctuation determinant. Thirdly and finally these instantons were
localised to an absolute scale which allowed the construction of a dilute gas.

We have succeeded, in the context of the pure-Skyrme $O(3)$ model, to calculate
the density of the (dilute) instanton gas \r{g7} explicitly, which would not
have been possible if our theory did not have an absolute scale. This
would not have been achieved in
practice if we could not have availed of explicit intantons and if
we had not calculated the regularised determinants explicitly. The
result \r{g7} allowed the calculation of the shift in vacuum energy
\r{g12} due to the instantons, a quantity which to our knowledge has not 
previously been calculated explicitly for a field theory, but only in 
quantum mechanics~\cite{GP}. In this sense, our results should be 
of interest, in addition to having achieved our main aim of 
tackling the semiclassical instanton analysis of
a theory with Skyrme-like kinetic terms.

\section*{Acknowledgements}

We are grateful to D. Diakonov, A. Sedrakyan and A. Morozov for enlightening discussions.
ST acknowledges the support of the Deutsche Forschungsgemeinschaft (DFG).
This work was partially supported by FORBAIRT (Ireland) under project
IC/98/035.

\end{document}